\documentclass[11pt,twocolumn]{article}

\usepackage[margin=1in]{geometry}
\usepackage[T1]{fontenc}
\usepackage[utf8]{inputenc}
\usepackage{textcomp}
\usepackage{tgheros}

\usepackage{amsmath,amssymb}
\usepackage{microtype}
\usepackage{graphicx}
\usepackage{booktabs}
\usepackage{array}
\usepackage{enumitem}
\usepackage{parskip}
\usepackage{xcolor}
\usepackage{caption}
\usepackage{titlesec}
\usepackage{float}
\usepackage{fancyhdr}
\usepackage{hyperref}

\definecolor{journalblue}{HTML}{1F4E79}
\definecolor{journalteal}{HTML}{0E7C86}
\definecolor{journalgray}{HTML}{4B5563}
\newcommand{\journaldate}{18 Sep 2025}

\setlist[itemize]{leftmargin=1.35em}
\setlist[enumerate]{leftmargin=1.7em}
\hypersetup{
  colorlinks=true,
  linkcolor=journalblue,
  citecolor=journalblue,
  urlcolor=journalteal
}

\titleformat{\section}{\Large\bfseries\color{journalblue}}{\thesection}{0.6em}{}
\titleformat{\subsection}{\large\bfseries\color{journalteal}}{\thesubsection}{0.6em}{}
\titlespacing*{\section}{0pt}{1.25em}{0.35em}
\titlespacing*{\subsection}{0pt}{0.95em}{0.3em}

\begin{document}

\twocolumn[
\begin{@twocolumnfalse}

\begin{center}
{\LARGE\bfseries\color{journalblue} Efficient and Accessible Discrete Choice Experiments: \\ The DCEtool Package for R}\par
\vspace{0.6em}
{\large\color{journalteal} Daniel P\'erez-Troncoso}\par
{\normalsize Department of Statistics and Modelling, Outcomes'10}\par
\vspace{0.35em}
{\small\textcolor{journalgray}{Manuscript date: \textbf{\journaldate}}}
\end{center}

\vspace{0.6em}
\begin{center}
{%
\setlength{\fboxsep}{9pt}
\fcolorbox{journalblue!45}{journalblue!5}{%
\begin{minipage}{0.92\linewidth}
\textbf{\textcolor{journalblue}{Highlights}}
\begin{itemize}[itemsep=0.25em,topsep=0.45em]
  \item DCEtool is a free R package with a Shiny interface that facilitates the design, implementation, and analysis of discrete choice experiments.
  \item It incorporates statistically efficient design algorithms and allows for decoding and labeling of the design matrix.
  \item The tool enables users to create local surveys, test them interactively, and estimate conditional and mixed logit models.
  \item DCEtool supports willingness-to-pay estimation and includes built-in support for serial DCEs.
  \item Its pedagogical interface makes it suitable for both researchers new to DCEs and those looking to streamline their workflow.
\end{itemize}
\end{minipage}%
}
}
\end{center}

\vspace{0.9em}
\begin{abstract}
Discrete Choice Experiments (DCEs) are widely used to elicit preferences for products or services by analyzing choices among alternatives described by their attributes. The quality of the insights obtained from a DCE heavily depends on the properties of its experimental design. While early DCEs often relied on linear criteria such as orthogonality, these approaches were later found to be inappropriate for discrete choice models, which are inherently non-linear. As a result, statistically efficient design methods, based on minimizing the D-error to reduce parameter variance, have become the standard. Although such methods are implemented in several commercial tools, researchers seeking free and accessible solutions often face limitations. This paper presents DCEtool, an R package with a Shiny-based graphical interface designed to support both novice and experienced users in constructing, decoding, and analyzing statistically efficient DCE designs. DCEtool facilitates the implementation of serial DCEs, offers flexible design settings, and enables rapid estimation of discrete choice models. By making advanced design techniques more accessible, DCEtool contributes to the broader adoption of rigorous experimental practices in choice modelling.
\end{abstract}

\vspace{0.65em}
\textbf{\textcolor{journalblue}{Keywords:}} discrete choice experiments, DCEtool, R, package, preference elicitation

\vspace{0.8em}
\end{@twocolumnfalse}
]

\section{Introduction}

Over the past two decades, the design of Discrete Choice Experiments (DCEs) has been the subject of intense debate. Traditionally, researchers relied on orthogonal designs to reduce full factorial structures \cite{rose2014}. However, it was soon pointed out that orthogonality is not suitable for DCEs due to the non-linearity of discrete choice models \cite{bliemer2010}.

To address this issue, many researchers began using statistically efficient designs. These designs aim to increase the precision of parameter estimates by minimizing the design's D-error \cite{bunch1996}. In essence, minimizing the D-error helps to reduce the standard errors of the parameters estimated from DCE data. Since this minimization involves an iterative process, software-based optimization routines are required. Examples include Ngene \cite{ngene2018}, the \texttt{choiceeff} macro for SAS \cite{saschoicEff}, the \texttt{dcreate} module for Stata \cite{hole2015}, and the \texttt{idefix} package for R \cite{traets2020}. However, until now, there has been no free software with a user-friendly interface that facilitates this process.

This article introduces DCEtool, an R package with a visual interface built with Shiny, which makes high-quality design techniques accessible to both novice and experienced researchers. DCEtool can generate and decode DCE design matrices, create local interactive surveys, and analyze responses using discrete choice models. It integrates:

\begin{itemize}
  \item code from the \texttt{idefix} package to construct experimental designs,
  \item code from the \texttt{survival} and \texttt{mlogit} packages to estimate models, and
  \item newly developed code to present a survey that can be answered live to test the design interactively.
\end{itemize}

Because DCEtool includes all the tools needed to run and analyze a DCE, researchers can generate a survey, complete it themselves, and estimate a discrete choice model within minutes. This makes DCEtool especially valuable as a pedagogical tool for those learning about DCEs. Furthermore, the software includes built-in functionality to implement both the serial DCE approach proposed by Bliemer and Rose \cite{bliemer2010} and the method proposed in Pérez-Troncoso \cite{perez2022}. Since serial designs are often time-consuming when implemented manually, DCEtool offers a practical and cost-free way to apply them.

\section{Requirements and Installation}

\subsection{Requirements}

The program has been tested on Windows, macOS, and Linux. To run properly, it requires both R \cite{rcore2025} and RStudio \cite{posit2025}. All other dependencies are automatically installed during the DCEtool installation process. There are no specific hardware requirements, but the program performs better on computers with faster CPUs and higher amounts of RAM.

\subsection{Installation}

DCEtool is available on CRAN, so it can be installed like any other package from the official repository:

\begin{enumerate}
  \item \texttt{install.packages("DCEtool")}
\end{enumerate}

Once installed, the graphical interface will appear after typing the following commands:

\begin{enumerate}[start=2]
  \item \texttt{library(DCEtool)}
  \item \texttt{DCEtool()}
\end{enumerate}

However, the most recent version of the software is often available earlier on GitHub (\url{https://github.com/danielpereztr/DCEtool}). To install it from GitHub in RStudio, use the following commands:

\begin{enumerate}[start=4]
  \item \texttt{install.packages("devtools")}
  \item \texttt{library(devtools)}
  \item \texttt{install\_github("danielpereztr/DCEtool")}
  \item \texttt{library(DCEtool)}
  \item \texttt{DCEtool()}
\end{enumerate}

Due to CRAN repository policies, packages hosted there can only be updated every 1--2 months. As a result, the GitHub version might contain more recent changes. Note that steps 4 to 8 may not work on some Linux systems. If that happens, a possible solution is provided in Section 4.1.

\section{Instructions}

The user interface (UI) is organized into five main tabs (in addition to a Home tab, which provides basic information about the software, and other tabs that may be added in future versions):

\begin{itemize}
  \item Design settings --- to configure the experimental design.
  \item Design matrix --- to display the generated design based on the user settings.
  \item Create a survey --- to build the survey interface.
  \item Survey --- to preview and interact with the survey locally.
  \item Results --- to combine survey responses with the design matrix and estimate discrete choice models.
\end{itemize}

The order of the tabs reflects the recommended logical and chronological workflow. Readers are encouraged to follow the sections in this order, replicating the steps and experimenting with changes to explore the software's capabilities.

\subsection{Design settings}

In this tab, users are asked to input all the settings required to build the experimental design. Before doing so, users must decide on the attributes and levels to be included. Table~\ref{tab:attributes-levels} provides an example with four attributes and 3, 2, 3, and 3 levels, respectively. A full factorial design based on these selections would result in $3 \times 2 \times 3 \times 3 = 54$ alternatives. If these alternatives were combined into pairs (e.g., choice sets with two alternatives), the total number of possible sets would be $(54 \times 53)/2 = 1431$, which is too many to be realistically used in a survey. Since 70\% of DCEs use between 3 and 7 attributes \cite{johnson2013}, most designs are a reduced subset of the full factorial design, that is, a selection of choice sets drawn from the full set of possibilities.

The selection of an optimal subset of choice sets can be carried out using exchange algorithms from the \texttt{idefix} package, which is integrated into DCEtool. Through the UI, users only need to specify:

\begin{itemize}
  \item the number of attributes and levels,
  \item the number of alternatives and choice sets,
  \item whether to include an opt-out alternative,
  \item whether to use a Bayesian design,
  \item a random seed, and
  \item a set of priors.
\end{itemize}

Whether to use a Bayesian design is up to the user and goes beyond the scope of this article. More details can be found in Kessels et al. \cite{kessels2011}. The random seed can be any number and ensures that the same design is obtained if the parameters and seed are repeated. The set of priors should reflect estimates from a pilot DCE. If no pilot has been conducted, a pilot design can still be created by setting all priors to zero, which is the default in DCEtool.

\begin{table}[H]
\centering
\caption{Attributes and levels selection.}
\label{tab:attributes-levels}
\begin{tabular}{p{0.34\linewidth}p{0.56\linewidth}}
\toprule
\textbf{Attribute} & \textbf{Levels} \\
\midrule
Effectiveness & 70\%, 80\%, 90\% \\
Required dosage & 1 dose, 2 doses \\
Adverse events & 1 in 1000 patients, 1 in 500 patients, 1 in 100 patients \\
Out-of-pocket cost & 100\texteuro, 150\texteuro, 200\texteuro \\
\bottomrule
\end{tabular}
\end{table}

\begin{figure}[H]
  \centering
  \includegraphics[width=\linewidth]{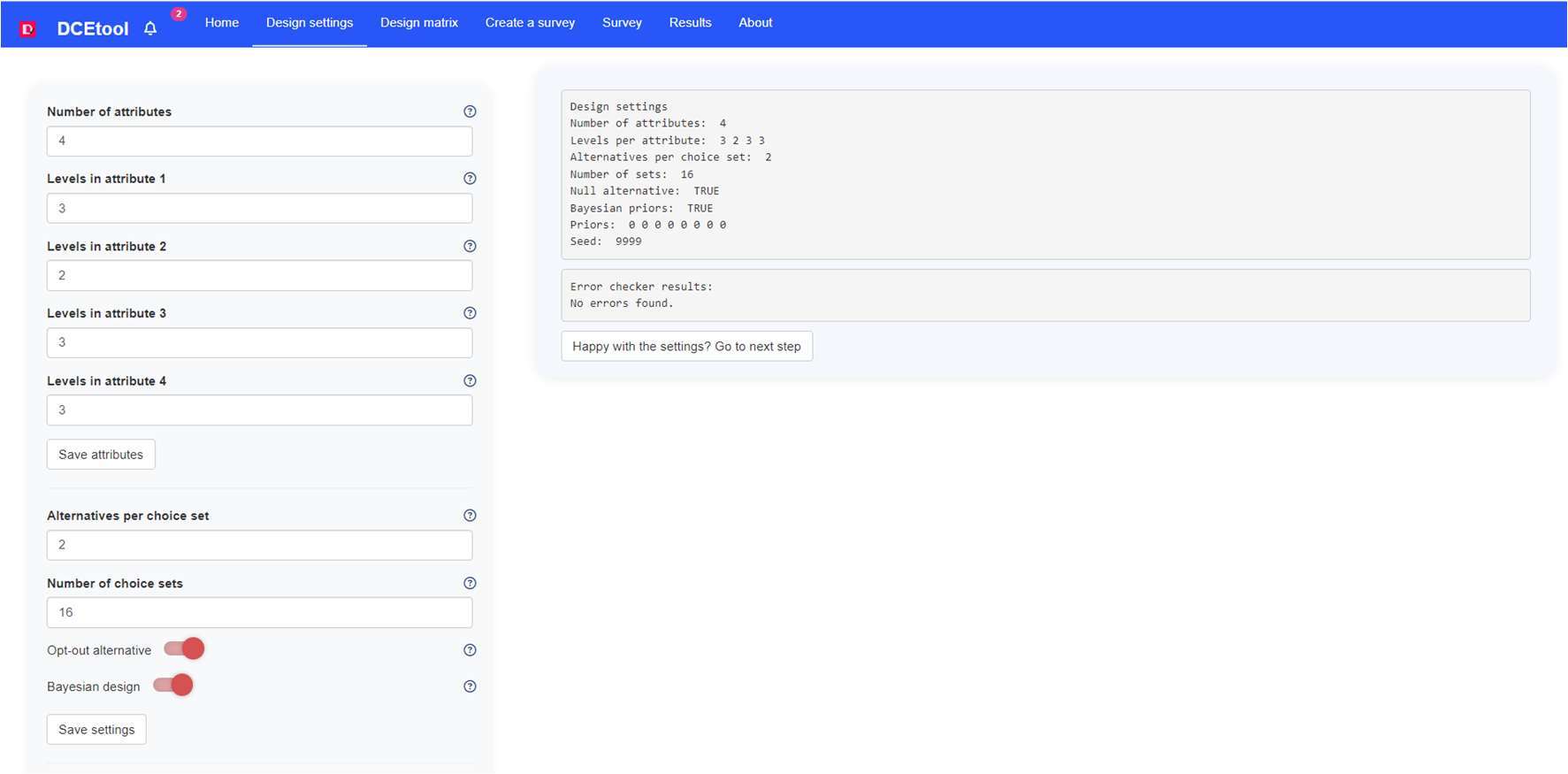}
  \caption{Design settings tab.}
  \label{fig:design-settings}
\end{figure}

Figure~\ref{fig:design-settings} displays the appearance of the Design settings tab after inputting the design specifications. If DCEtool detects incompatible settings (e.g., too few sets given the number of attributes and levels), an error message will appear. Once all settings are valid, users can proceed by clicking the \emph{Go to next step} button.

\subsection{Design matrix}

Once in the Design matrix tab, an efficient experimental design will be generated when the user clicks on \emph{Generate design}. The computation time may vary depending on the computer's CPU and RAM performance. After a few seconds or minutes (depending on the design's size) and once the loading animation ends, the design matrix will be displayed as a table.

At this stage, the matrix might be difficult to interpret for many users. To address this, DCEtool includes a decoding function. To use it, the user must first label the attributes and levels. This is done by clicking on \emph{Name the attributes}, entering the name of each attribute, and clicking \emph{Save names}. Following the example in Table~\ref{tab:attributes-levels}, the attribute names would be Effectiveness, Required dosage, Adverse events, and Out-of-pocket cost.

Next, under \emph{Change the level names}, the user should input the names of each level for each attribute. For example, for Effectiveness, the levels would be 70\%, 80\%, and 90\%. After entering these, the user should click \emph{Save level 1}, then continue with the levels of the next attribute (e.g., Required dosage).

Once all names have been provided, the user clicks on \emph{Change names in the design matrix} to apply the labels. The labelled design can then be saved as an Excel file by clicking \emph{Save design}. If the user already has a DCEtool-generated Excel file, they can skip the previous steps and upload it using the \emph{Browse} button.

\begin{figure}[H]
  \centering
  \includegraphics[width=\linewidth]{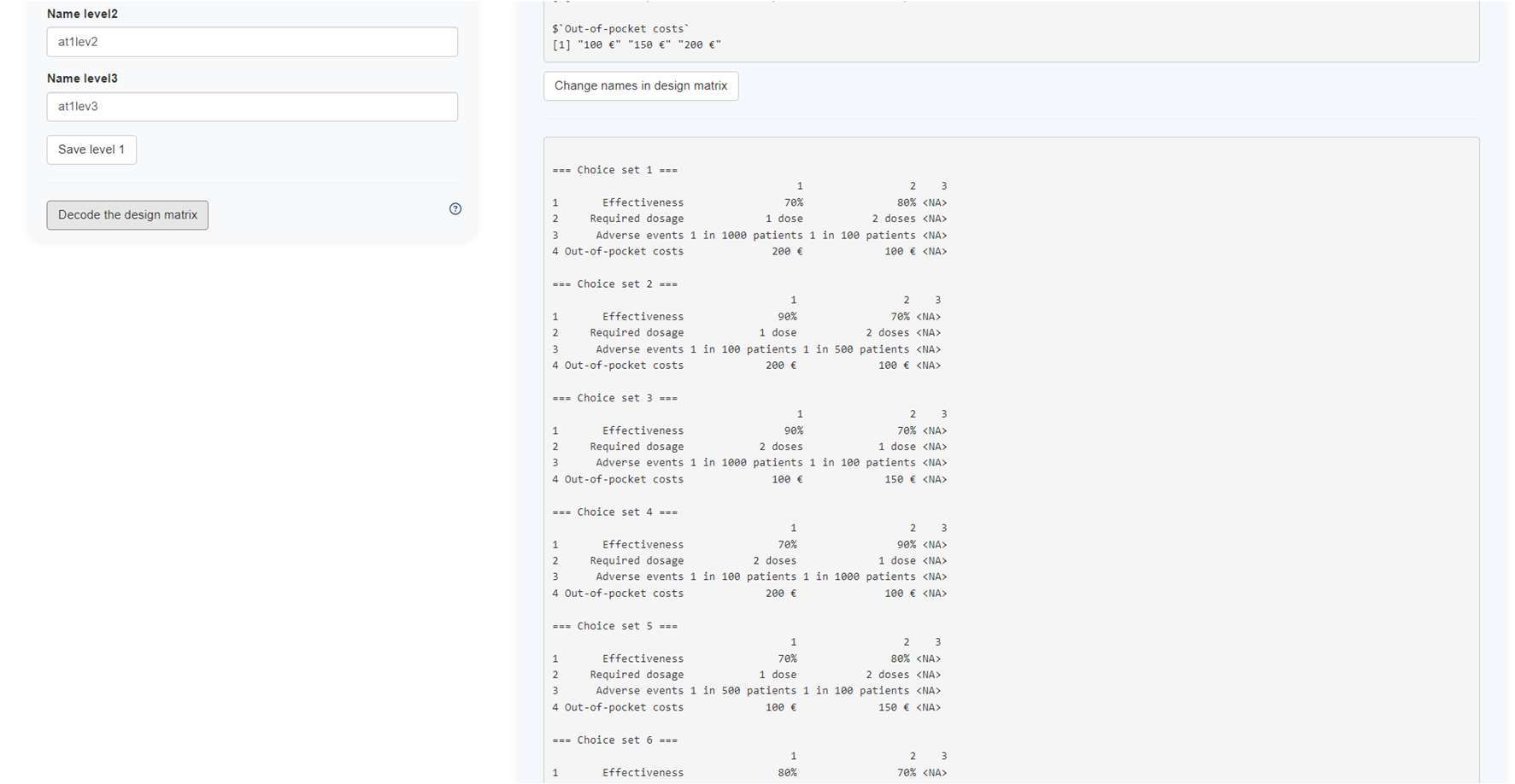}
  \caption{DCEtool output after clicking on \emph{Decode the design matrix}.}
  \label{fig:decode-matrix}
\end{figure}

Finally, by clicking \emph{Decode the design matrix}, the choice sets will be displayed in plain text (see Figure~\ref{fig:decode-matrix}). This output can be exported and used to build a paper-based survey or uploaded into another online survey platform.

\subsection{Create a survey}

After decoding the design matrix in the previous section, users can test their survey directly within DCEtool by responding to it and analyzing the results. To do this, the \emph{Create a Survey} tab allows users to add both an introductory and a final text to the survey, as well as a personalized label for each choice set.

The introductory and final texts can be written using Markdown syntax, allowing for basic formatting. The alternative labels can be customized freely; however, since DCEtool is primarily designed for unlabeled DCEs, a common option is to label the alternatives as Option 1, Option 2, etc. If an opt-out alternative was included in the design, it can be labeled accordingly, for example, as Opt-out.

\begin{figure}[H]
  \centering
  \includegraphics[width=\linewidth]{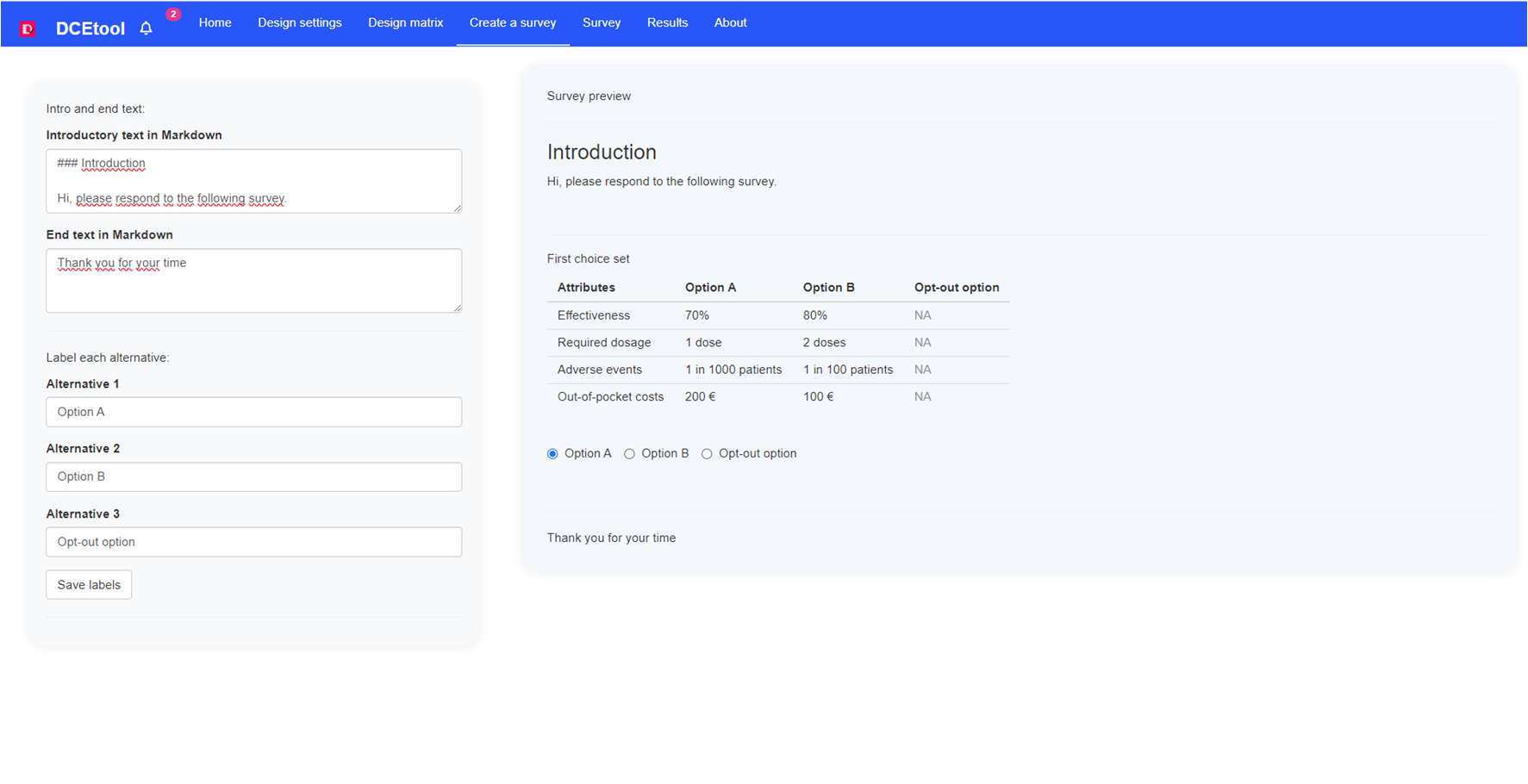}
  \caption{\emph{Create a survey} tab preview.}
  \label{fig:survey-preview}
\end{figure}

Once the labels have been saved, a preview of the survey will be displayed (see Figure~\ref{fig:survey-preview}), allowing the user to review the structure and content before proceeding.

\subsection{Survey}

Once in the Create a Survey tab, the user will be presented with three options regarding the survey mode. Selecting \emph{No} means that a serial approach will not be applied, and all respondents will complete the same version of the survey.

The \emph{Bliemer \& Rose (each respondent)} option activates the serial strategy proposed by Bliemer and Rose \cite{bliemer2010}, in which a new DCE design is generated after each individual response, to be shown to the next respondent. This option is only available when responding to the survey directly in DCEtool, as it requires the design matrix to be updated in real time after each answer.

The third option, \emph{Each 5 respondents}, is an extension of the Bliemer and Rose approach proposed in Pérez-Troncoso \cite{perez2022}, designed to reduce computational demands by updating the design every five responses instead of after each one.

Both serial modes are particularly useful for pedagogical purposes. However, in the current version of the app, they can only be used locally on the same machine. If the goal is to test the DCE as it will be used in the actual study, i.e., collecting multiple responses to the same survey, it is recommended to avoid the serial mode.

\begin{figure}[H]
  \centering
  \includegraphics[width=\linewidth]{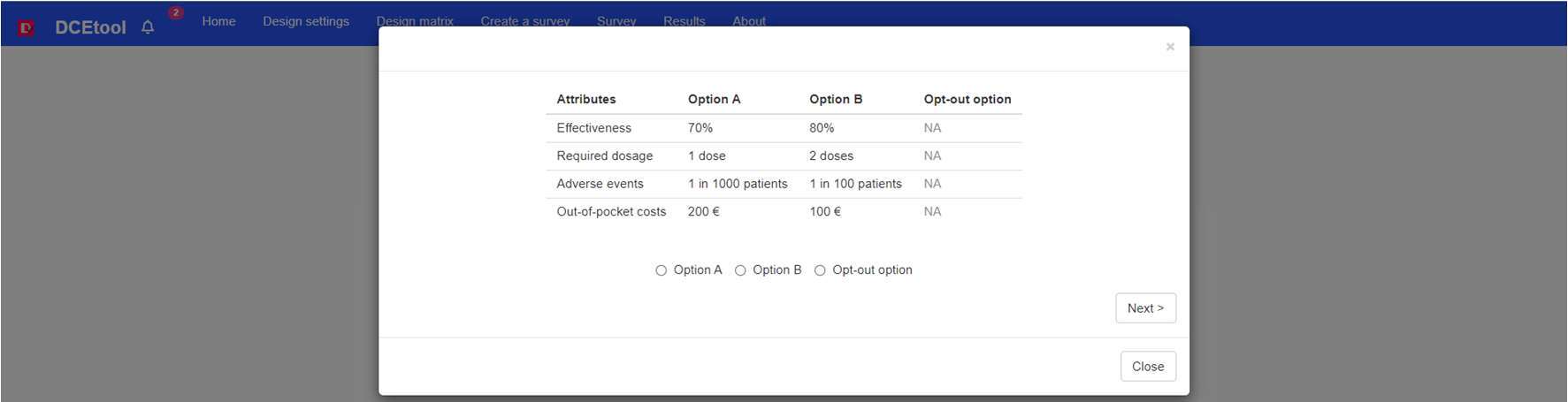}
  \caption{Choice set in the survey.}
  \label{fig:choice-set}
\end{figure}

After starting the survey, the user can respond to the DCE as many times as desired. When ready to analyze the collected responses, the user should be on the final page of the survey (where the final text is displayed). Instead of clicking \emph{Next respondent}, they should click \emph{Close} and then navigate to the \emph{Results} tab.

\subsection{Results}

In the Results tab, the user will find a table containing the collected responses, already coded and ready to be analyzed using DCEtool or exported to other software. If the estimation is to be performed outside DCEtool, the results can be saved as an Excel file by clicking the \emph{Save results} button.

DCEtool supports the estimation of conditional and mixed logit models, as well as willingness-to-pay (WTP) calculations. First, in the \emph{Data} section, the user can recode the price variable as a continuous variable to enable WTP estimation (only applicable if a price or cost attribute was included in the DCE). To do this, the user must check \emph{Code price as continuous variable}, select the levels corresponding to the price attribute (e.g., 150\texteuro{} and 200\texteuro{} in our example), and input their numerical values (e.g., 100 --- omitted level ---, 150, and 200).

Once this variable has been added to the dataset, the user can proceed to the \emph{Estimation} section to estimate the conditional logit model. The model specification is straightforward: \emph{choice} must be selected as the dependent variable, the remaining levels as independent variables (if the price variable was coded as continuous, it will appear as \texttt{cont\_price}, replacing all price levels), and \texttt{gid} as the group identifier.

\begin{figure}[H]
  \centering
  \includegraphics[width=\linewidth]{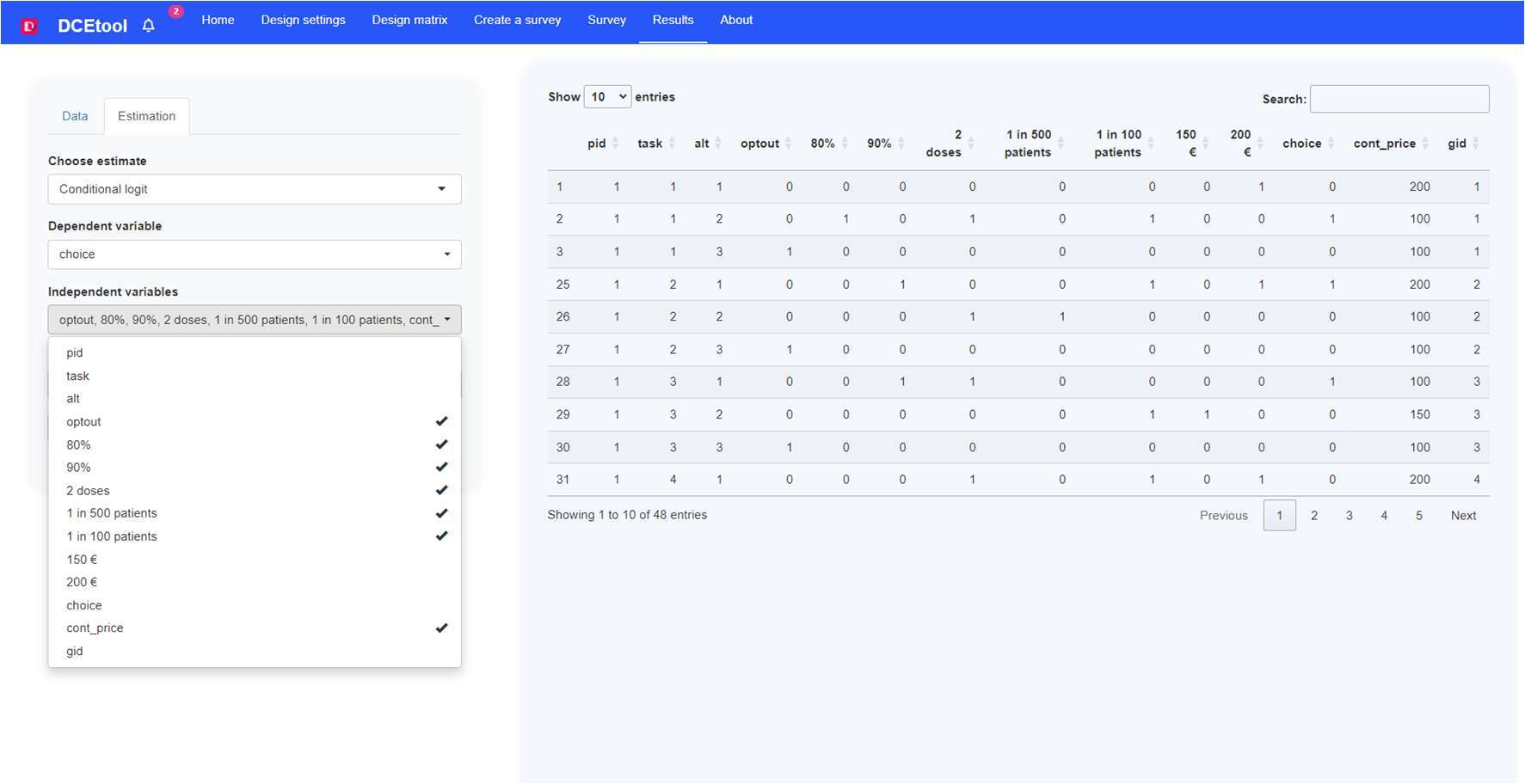}
  \caption{Estimation section in Results.}
  \label{fig:estimation-section}
\end{figure}

\begin{figure}[H]
  \centering
  \includegraphics[width=\linewidth]{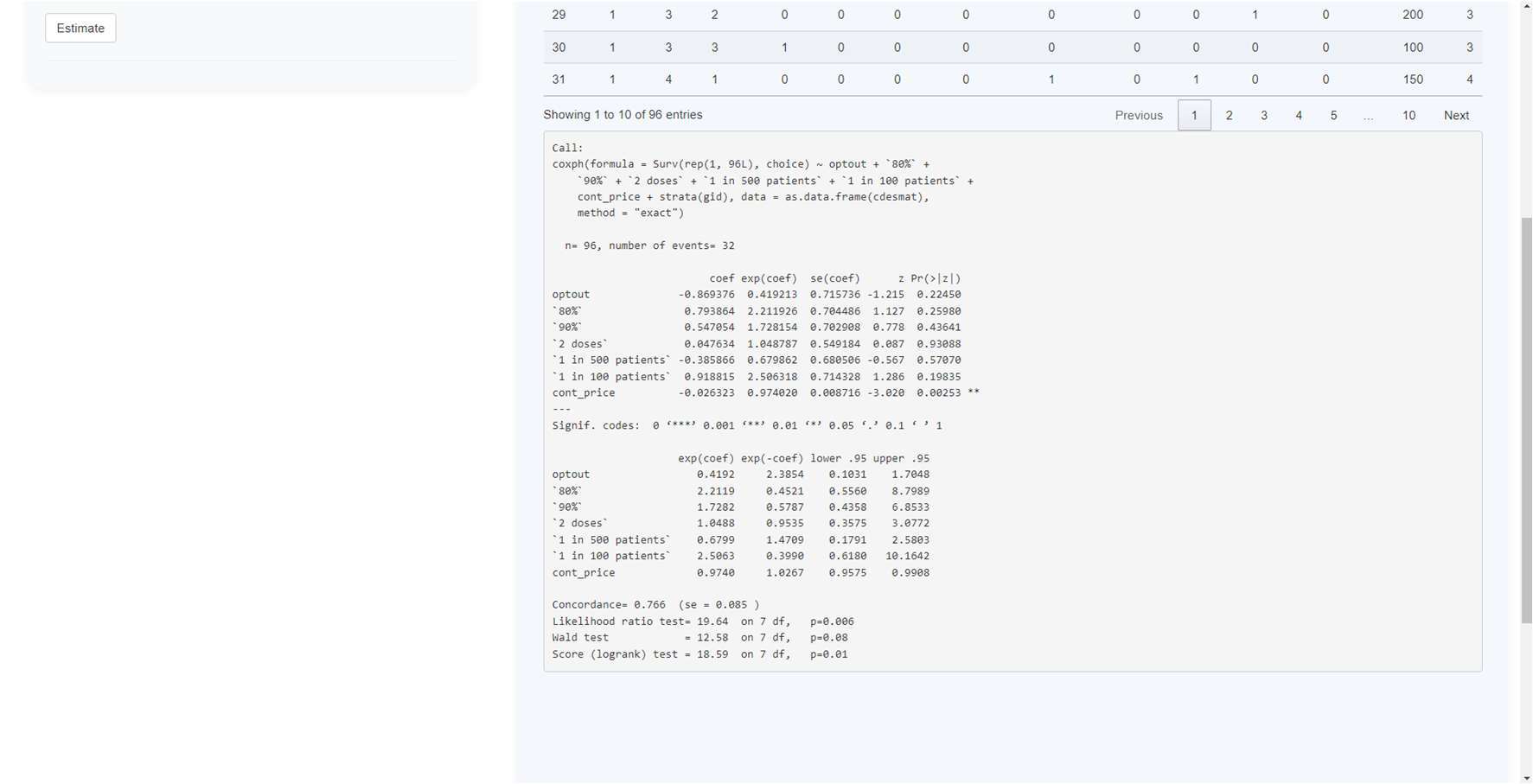}
  \caption{Results tab.}
  \label{fig:results-tab}
\end{figure}

After clicking the \emph{Estimate} button, a results box with the model output will appear below the data table (see Figure~\ref{fig:results-tab}).

\textbf{Note:} The coefficients and p-values shown in the illustrated examples do not have sufficient significance (or logical meaning), since they were obtained by responding semi-randomly to the survey in order to test functionality.

After estimating the logit model, the user can compute the numerical values of willingness to pay (WTP) by selecting \emph{Willingness to pay} from the drop-down menu. Then, the user must specify the price variable (e.g., \texttt{cont\_price}) and select the attribute levels of interest (e.g., 80\%, 90\%, 2 doses, 1 in 500 patients, 1 in 1000 patients).

It is important to note that base levels (e.g., 70\%, 1 dose, and 1 in 100 patients) will not appear as options. This is because WTP is interpreted as the additional amount of money the average respondent is willing to pay to receive the benefit associated with a particular level relative to its base level.

Finally, if the user selects \emph{Figures} from the drop-down menu, a graphical representation of the coefficients and their confidence intervals will be displayed.

\section{Conclusions}

The rejection of traditional methods (such as orthogonal designs) due to their inadequacy for non-linear discrete choice models has led to the development of statistically efficient design criteria. This methodological evolution has been driven by the need to improve the precision and quality of insights derived from DCEs. However, the absence of free, user-friendly software has long posed a barrier to wider adoption. DCEtool was developed to address this gap.

DCEtool enables the creation, decoding, and analysis of discrete choice experiments with robust design properties through a Shiny-based visual interface in R. It integrates best practices from established packages while introducing new features for survey presentation and model estimation. Notably, it simplifies the implementation of serial discrete choice experiments, which can enhance the precision of parameter estimates.

With a pedagogical orientation that supports beginners and accelerates the workflow of experienced users, DCEtool is a valuable addition to the researcher's toolkit. Its ability to dynamically adjust designs and incorporate Bayesian options adds to its flexibility. By streamlining the process of generating choice sets, running surveys, and estimating models, DCEtool contributes to more accurate and efficient research practices and advances the field of discrete choice modelling.

\end{document}